\documentclass[11pt]{article}
\usepackage{mor,epsfig}

\def\Journal#1#2#3#4{{#1} {\bf #2}, #3 (#4)}

\def\NIM{\em Nucl. Instrum. Methods}

\def\PRL{\em Phys. Rev. Lett.}
\def\PRD{{\em Phys. Rev.} }

\def\MPLA{{\em Mod. Phys. Lett.} A}
\def\PR{\em Phys. Rep.} 

\def\be{\begin{equation}}
\def\ee{\end{equation}}
\def\bea{\begin{eqnarray}}
\def\eea{\end{eqnarray}}

\newcommand{\ttbar}{ {\rm t} \bar{{\rm t}} }

\newcommand{\ppbar}{ {\rm p} \bar{{\rm p}} }

\newcommand{\et}{{\rm E}_{\scriptscriptstyle\rm T}}
\newcommand{\met}{\mbox{$\protect \raisebox{.3ex}{$\not$}\et$}}

\newcommand{\lqlqbar}{ {\rm LQ} \bar{{\rm LQ}} }

\begin{document}
\vspace*{4.cm}
\title{RECENT CDF AND  D\O~ RUN~I RESULTS}

\author{ G.V. VELEV}

\address{Fermi National Accelerator Laboratory,Batavia, IL 60510, USA\\
 (on behalf of  the CDF and  D\O~ Collaborations)}

\maketitle\abstracts{
We  summarize   some of the the most recent  CDF and  D\O~ results from  
the 1992-1995 collider run at the Fermilab Tevatron. These include a  detailed 
examination of the heavy flavor content of W+jet data made by CDF. 
We found in this study that the rate and the kinematic properties 
of the event subsample, featuring soft lepton and secondary vertex in the same jet,
are statistically difficult to accommodate with the Standard Model  simulation.
CDF has also searched for  new physics in events with a  photon, a lepton and $\met$. 
Finally,  the  results of the two collaborations in their search for the 
first, second and third  generations leptoquarks are presented.}

\section{Introduction}
In this paper\footnote[0]{Presented at XXXVII Rencontres de Moriond, EW Interactions and Unified Theories, LesArcs 1800, France March 9-16, 2002}, we  describe some of the most recent results 
obtained by the  CDF and  D\O~ collaborations in the search for physics 
beyond the Standard Model.  
Both experiments operate at the Tevatron $\ppbar$ collider at Fermilab. The 
results are based on the  analysis of  data samples collected during the 
1992-1995 run
known as Tevatron Run~I, and  are based on an integrated 
luminosity exceeding  100 $pb^{-1}$ per detector.

The CDF and  D\O~  Run~I detectors are  described  in Ref.~\cite{cdf} and Ref.~\cite{d0} respectively.

\section{CDF detailed examination of the heavy flavor content of W+jet data}
A study of the  properties of events containing a W boson and associated jets (W+jet sample),
provides a good opportunity to test our understanding of the standard model (SM) QCD 
and electroweak predictions. This event sample was used to establish the top quark 
discovery~\cite{discovery}. The whole sample was also  exploited to perform a  measurement of 
the top quark mass~\cite{massprd} and 
\hglue1truept{\hskip3.0in{
\parbox {8.2cm} {
of its  production cross section~\cite{xsecprd} after assuming that any observed
excess of beauty-tagged data was due to $\ttbar$ production. In this study 
a complementary approach is adopted~\cite{superjet}.
We use the theoretical estimate  of $\sigma_{\ttbar}$  and we  test the  compatibility of the SM prediction 
with the observed number of 
different tags~\footnotemark   as a function of jet multiplicity.
}} \footnotetext {CDF uses two different methods for identifying (tagging) heavy 
quark jets~\cite{discovery}.
The first technique uses the silicon microvertex detector (SVX) to locate secondary vertices from 
b and c-hadron decays (SECVTX tag). The second one (SLT) searches for a relatively soft lepton ($e$ or $\mu$)
contained in the jet cone and produced  by  these semileptonic decays.}

\begin{figure}
\unitlength1cm
\begin{minipage}[p]{7.0cm}
\psfig{figure=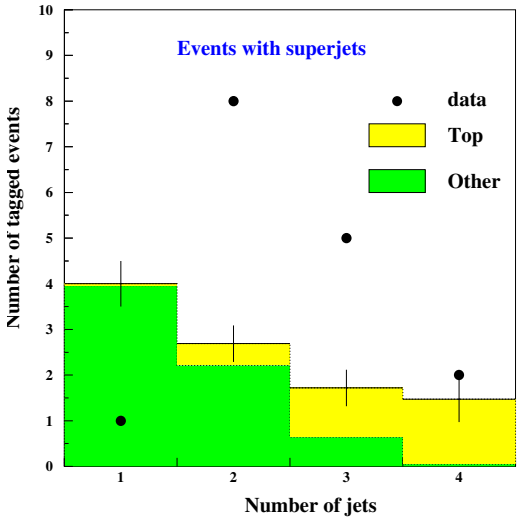,height=3.0in}
\caption{Comparison between observed and predicted number of W+jet events with a supertag as a function of the
number of jets in the event. 
\label{fig:superjet}}
\end{minipage}
\end{figure}

\hfill
\begin{table}
\vglue1truept\vskip-3.65in{\hglue1truept{\hskip2.80in{
\begin{tiny}
\begin{minipage}[p]{8.0cm}
\begin{tabular}{|ccccc|}
\hline\hline 
        & W+1jet & W+2jets & W+3jets & W+$\ge$4jets  \\
\hline
\multicolumn{5}{|c|}{SECVTX events}\\
SM (ST) & $64.4\pm6.5$  & $29.6\pm2.7$ & $12.9\pm1.9$ & $8.9\pm2.0$\\
SM (DT) &                 & $ 2.4\pm0.6$ & $ 3.2\pm0.8$ & $4.0\pm1.0$\\
\hline
Data (ST) &    66           & 35 & 10 & 11 \\
Data (DT) &                 &  5 &  6 &  2 \\
\hline
\multicolumn{5}{|c|}{SLT events} \\
SM (ST) & $137.75\pm11.29$  & $46.1\pm5.7$ & $12.9\pm1.9$ & $8.9\pm2.0$\\
SM (DT) &                 & $ 0.1\pm0.1$ & $ 0.1\pm0.1$ & $0.2\pm0.1$\\
\hline
Data (ST) &    146          & 56 & 17 & 8  \\
Data (DT) &                 &  0 &  0 &  0 \\
\hline
\multicolumn{5}{|c|}{ Superjet events} \\
SM (SJ) & $4.00  \pm0.50 $  & $ 2.7\pm0.4$ & $ 1.7\pm0.4$ & $1.5\pm0.5$\\
SM (DT) &                 & $ 0.3\pm0.1$ & $ 0.4\pm0.1$ & $0.5\pm0.1$\\
\hline
Data (SJ) &    1            & 8  &  5 & 2  \\
Data (DT) &                 &  2 &  3 &  0 \\
\hline
\hline 
\end{tabular}
\caption{Observed and  predicted number of W events with SECVTX tag (top lines), soft lepton tag (center lines) and the events with a supertag (bottom part). 
\label{tab:svx-slt}
}
\end{minipage}
\end{tiny}
}}}

\vspace{-0.0in}
\end{table}  

\vspace{-0.5cm}
The top part of Table~\ref{tab:svx-slt} summarizes the number of observed and predicted W events 
with one (ST) or (DT) two SECVTX tags in the accompanying jets. The comparison for 
SLT tags is  shown at the center. 
The probability that the observed numbers of events with at least one SECVTX (SLT) tag 
are consistent with the prediction in all jet bins is 80\% (56\%)~\cite{superjet}.

Studying the correlation between the taggers we found that the SM simulation does not 
predict well the  number of events with a SLT and a SECVTX  tag in the same jet. 
We called these tags and jets supertags and superjets, respectively.    
The  comparison between the  observed number of events with supertag and the SM prediction 
is  summarized in the  bottom lines of Table~\ref{tab:svx-slt} and is shown in Fig.~\ref{fig:superjet}
 The probability that  the observed numbers 
of events with at least one superjet fluctuate to the prediction in all four jet bins 
is 0.4\%. The  $a~posteriori$ probability of observing no less then 13 in the W+2,3 jet bins,
where $4.4\pm0.6$ are expected from SM sources, is 0.1\%.

We selected a complementary data sample which have a close composition to the superjet sample.
This sample consists of  W+2,3 jet events with a SECVTX tag, but no SLT tag in the same jet. However,
we  require  that at least one of the SECVTX tagged jets contains a soft lepton candidate 
track. After all of requirements we left with 42 W+2,3 jet events, while $41.2\pm3.1$ events are expected 
from the SM processes.

When we examined the additional jets in the superjet sample  we found 5 events with an 
additional  SECVTX tag. If the 13 events are a fluctuation from SM processes, we can 
expect to find $1.8\pm0.3$ events with double SECVTX tag. 
Taking into account the high probability of finding a SECVTX tag in the additional jets, we decided 
to name them conventionally as b-jets. 
             
If the 13 events are a statistical fluctuation, their kinematics would be consistent 
with the SM simulation and with the kinematics of the complementary sample. We choose two sets of 
9 variables to compare the samples~\cite{superjet}. The~~first set includes:~~$d\sigma/dp_{T}$~~and 
$d\sigma/d\eta$ of~ the lepton, 

\hglue1truept{\hskip2.3in{
\parbox {9.4cm} {
superjet and additional ``b-jet(s)''~\footnotemark; the 
transverse energy and the rapidity of the system ($\ell+suj+b$), which is strongly correlated 
with the missing transverse energy and the rapidity of the object producing $\met$; and 
the azimuthal angle $\phi^{\ell,suj+b}$ between the primary lepton and 
the superjet-additional jet(s) (b-jet(s)) system. These 9 variables are sufficient to 
describe the kinematics of the  final state with relatively modest correlations. 

\hskip0.6cm A second set of  9 variables was also tried, including: 
the corrected transverse 
missing energy ($\met$); the W transverse mass ($M_{T}^{W}$); the invariant mass, rapidity,
and transverse energy  of the system $suj+b$
 ($M^{suj+b}$,$y^{suj+b}$ and $E^{suj+b}_{T}$);
the  invariant mass of the system $l+suj+b$
 }
}\footnotetext{In the sub-sample of 
W+3jet events the same variables with two entries per event are used.}

\begin{figure}
\unitlength1cm
\begin{minipage}[p]{7.0cm}
\psfig{figure=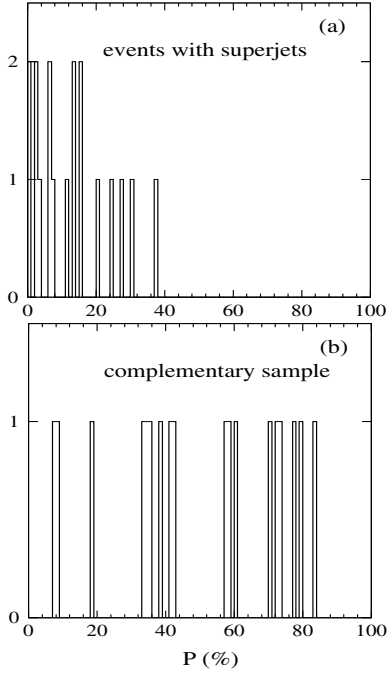,height=3.5in,width=2.0in}
\parbox[t]{2.0in}{\caption{
Distribution of the probabilities $\mathcal{P}$ that the 13 superjet events~(a) and 
the complementary sample~(b) are consistent with SM prediction.
\label{fig:KS-dist}}
}
\end{minipage}
\end{figure}

\hfill
\begin{table}
\vglue1truept\vskip-4.7in{\hglue1truept{\hskip2.5in{
\begin{tiny}
\begin{minipage}[p]{8.0cm}
\begin{tabular}{|ccccc|}
\hline\hline 
 Variable        &  \multicolumn{2}{c}{Events with superjet} & \multicolumn{2}{c|}{Complementary sample}\\
                 &  $\delta^{0}$ &  $\mathcal{P}$ (\%) & $\delta^{0}$ &  $\mathcal{P}$ (\%) \\ 
\hline
$E_{T}^{\ell}$ ($\met$)                      & 0.47(0.31)& 2.6(27.1)&  0.14(0.14) & 70.9(57.1) \\  
$\eta^{\ell}$ ($M_{T}^{W}$)                  & 0.54(0.36)& 0.1(13.1)&  0.12(0.16) & 72.7(38.2) \\
$E_{T}^{suj}$ ($M^{suj+b}$)                  & 0.38(0.36)& 11.1(4.0)& 0.15(0.12) & 43.0(58.9) \\
$\eta^{suj}$ ($y^{suj+b}$)                  & 0.36(0.35)& 15.2(7.1)& 0.13(0.14) & 73.4(34.9) \\
$E_{T}^{b}$($E^{suj+b}_{T}$)                 & 0.36(0.28)&  6.7(24.0)& 0.18(0.10) &  8.6(60.1) \\
$\eta^{b}$ ($M^{l+suj+b}$)                  & 0.38(0.31)&  6.8(21.0)& 0.11(0.15) & 80.0(33.6) \\
$E_{T}^{\ell+suj+b}$ ($\theta^{suj,b}$)      & 0.39(0.26)&  2.5(30.1)& 0.17(0.15) & 18.8(41.1) \\
$E^{\ell+suj+b}$ ($\phi^{suj,b}$)            & 0.31(0.31)& 13.8(15.3)& 0.19(0.10) &  7.8(83.8) \\
$\phi^{\ell,suj+b}$ ($\theta^{\ell,suj+b}$)  & 0 43(0.25)&  1.0(37.3)& 0.12(0.16) &  77.9(35.7) \\ 
\hline
\hline 
\end{tabular}
\caption{Summary of the KS comparison between data and simulation. The results for the two 
sets of kinematical variables are presented. 
\label{tab:KS-table}
}
\end{minipage}
\end{tiny}
}}}
\end{table}  

\vspace{-0.5cm}
\hskip-0.6cm 
 ($M^{l+suj+b}$);  the  angle and the azimuthal angle 
between the superjet and b-jets ($\theta^{suj,b}$,$\phi^{suj,b}$); and  the angle between the primary 
lepton and the $suj+b$ system ($\theta^{\ell,suj+b}$). 

The data of the  superjet and of the complementary samples are compared to SM 
montecarlo distributions using a 
Kolmogorov-Smirnov (KS) test. The Kuiper's definition of the test was applied: 
$\delta=max(F(x_{i})-H(x_{i}))+max(H(x_{i})-F(x_{i}))$. For each variable, the probability 
distribution of the KS  distance $\delta$ is determined with the ensemble of montecarlo experiments. 
In each montecarlo experiment, temporary templates are constructed. They account for the Poisson 
fluctuations in the number of events in the SM processes 
and  for the Gaussian uncertainties in the SM cross sections.  
From these temporary templates, we randomly generated a distribution with the same number of 
entries as in the data and evaluated the KS distance with respect to the nominal SM template.

The results of the KS comparison between data and simulation for the 18 kinematic 
distributions are  summarized in Table~\ref{tab:KS-table} 
and  presented in Fig~\ref{fig:KS-dist}. The complementary 
sample probabilities 
are flatly  distributed which indicates consistency with the SM simulation. On the other hand, one can
 notice that the 
distribution for the superjet events  cluster at low probabilities. This indicates the  
difficulty of the SM simulation to describe the kinematics of the supperjet events. Additional studies, 
combining all 18 probabilities, and determining the statistical significance of the 
observed discrepancies, were done in the Ref.~\cite{combined}.   
  
An extensive study of the properties of the superjets and/or superjet events was 
done~\cite{superjet}. This 
includes a detailed examination of the soft lepton tag parameters, a
check of the primary lepton isolation and lifetime, an investigation of the superjet properties 
(lifetime, hadronization) using  generic QCD data and SM montecarlo.
In addition, a number of background and acceptance studies were performed and no anomalous behavior
was found.  
However, removing the second 
level of the primary muon trigger 
requirement and extending the acceptance for the primary electrons 
up to $|\eta|<$1.5~\footnote{This was done by including the electrons not only from the 
central calorimeter $|\eta|<$1.0\, but from the CDF plug calorimeter too.},
4 additional W+2,3 jet events with superjets were found, while $0.42\pm0.06$ are 
expected from the SM.

\section{Search for a new physics in events with a photon and a lepton}
The inclusive production of a photon and  a lepton (e or $\mu$+$\gamma$+X) at large $P_{T}$ provides the 
opportunity to test many SM  predictions. Indeed, the  observation of an anomalous 
production rate of these events, possibly associated with additional photons, leptons, 
and  large missing transverse energy ($\met$) would be a clear indication of new 
processes beyond SM.

The interest to these events  originates from the appearance of the ee$\gamma\gamma\met$ 
event recorded by CDF. A supersymmetric model~\cite{Kane} designed to explain this event 
predicts the  production of  photons  from the radiative neutralino decay and of leptons from
the  chargino decay featuring $ \ell\gamma\met$ events as a signal. In addition, these
events can be explained by resonant smuon production with a single dominant R-parity 
violating coupling, in a model where gravitino is the lightest supersymmetric 
particle~\cite{Allanach}.   

Inclusive photon-lepton events are selected by CDF by requiring an isolated~\cite{gamma-lepton} 
central photon and lepton with $\et^{\gamma}>25$~GeV and $\et^{e,\mu}>25$~GeV respectively. 
The selection criteria for lepton and photons identification are described in detail in 
Ref.~\cite{selection}. A total of 77 events satisfied this selection: 29 photon-muon and 
48 photon-electron candidates. Figure~\ref{fig:radish} summarizes the selections criteria and 
shows the breakdown of the inclusive sample into the final categories. 

Rather than  looking for the specific characteristics of the events, like we did in 
the case  of  superjet events, here we simply compare the number of observed events 
to expected events. Without a specific model and assuming that our null hypothesis 
(the SM)  is correct, the significance of an observed excess is calculated as a 
probability $\mathcal{P}$ to obtain at least  the observed number of  events (N$_{0}$).
This probability $\mathcal{P}$(N$\ge$N$_{0}$/$\mu_{SM}$) is computed  from a large  
ensemble of montecarlo experiments in which each quantity used in the determination of  SM 
photon-electrons contribution is varied randomly with a Gaussian distribution, and the 
resulting event number is fluctuated according the Poisson law. The fraction of cases 
in the ensemble when N$\ge$N$_{0}$ gives $\mathcal{P}$(N$\ge$N$_{0}$/$\mu_{SM}$). 

The predicted and observed number of events for  two-body and multi-body 
selection are compared in Table~\ref{tab:exp}~\cite{gamma-lepton}. The most 
important SM contributers are  Z$\gamma$ production, where one of the Z-decay leptons escaped 
the detector, W$\gamma$ production, misidentified jets and electrons.  

The most significant deviation from the SM is observed in $\ell\gamma\met$X sample (it is 
outlined with bold box on Fig.~\ref{fig:radish}.) where 16 events are detected and $7.6\pm0.7$ 
are expected from SM sources. The {\it{a priori}} probability of  observing no less than 16 is 0.7\%, 
equivalent to 2.7$\sigma$ for a Gaussian distribution. 
      
\section{CDF and  D\O~ leptoquark searches}

Many extensions, for example see Refs.~\cite{leptotheory-seeD0-1,scalar-seeD0-2,leptotheory-seeD0-3},
 of the SM predict the existence of the 
leptoquarks (LQ), hypothetical color-triplet bosons with fractional electric charge that  
couple directly to leptons and quarks. Their masses are not  predicted from the models. 
They are assumed to be pair-produced at the Tevatron through a virtual gluon in the process 
$\ppbar\rightarrow g+X \rightarrow \lqlqbar+X$. For the scalar~\cite{scalar-seeD0-2} LQ, 
a production cross section is independent of the coupling between leptoquark, lepton and quark. 
In case of  vector~\cite{leptotheory-seeD0-1} LQ, the two specific possibilities of the coupling 
are~ assumed~ which~ result in minimal vector coupling (MV)  and  Yang-Mills coupling 

\hglue1truept{\hskip2.6in {
\parbox {8.5cm} {
(YM). In most models LQ are expected to couple only within a single generation because of 
the experimental limit imposed by the non-observation of flavor-changing neutral currents. 

\hskip0.6cm 
Using the Run~1 Tevatron~~ data, CDF~ and D\O~ have~ performed~ an extensive search for 
pair
}}

\begin{figure}
\unitlength1cm
\begin{minipage}[p]{7.0cm}
\psfig{figure=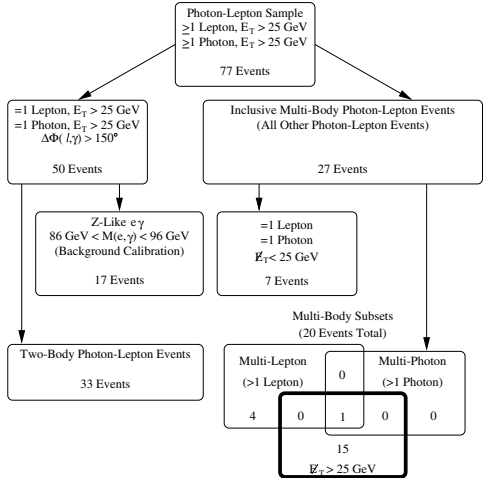,height=2.5in}
\caption{The diagram summarizes the selection criteria for the inclusive photon-lepton subsets. The bold box represents the sample with the largest 
deviation from SM prediction. 
\label{fig:radish}}
\end{minipage}
\end{figure}

\hfill
\begin{table}
\vglue1truept\vskip-3.4in{\hglue1truept{\hskip2.8in{
\begin{tiny}
\begin{minipage}[p]{8.0cm}
\begin{tabular}{|ccccc|}
\hline\hline 
Process & Two-Body& \multicolumn{3}{c|}{Multi-body} \\
        &  $\ell\gamma$X & $\ell\gamma$X & $\ell\gamma\met$X &$\ell\ell\gamma$X      \\
\hline
W+$\gamma$ & $2.7\pm0.3$  & $5.0\pm0.6$ & $3.9\pm0.5$ & - \\
Z+$\gamma$ & $12.5\pm1.2$ & $9.6\pm0.9$ & $1.3\pm0.2$ & $5.5\pm0.6$ \\
$\ell$+jet,jet$\rightarrow\gamma$ & $3.3\pm0.7$ & $3.2\pm0.6$ & $2.1\pm0.4$ & $0.3\pm0.1$ \\
Z$\rightarrow$ee,e$\rightarrow\gamma$ & $ 4.1\pm1.1$ & $1.7\pm0.5$ & $0.1\pm0.1$ & - \\
hadron+$\gamma$ & $1.4\pm0.7$  & $0.5\pm0.3$ & $0.2\pm0.1$ & - \\
$\pi$/K decay+$\gamma$ & $0.8\pm0.9$  & $0.3\pm0.3$ & $0.1\pm0.1$ & - \\
b/c  decay+$\gamma$ & $0.1\pm0.1$  & $<0.01$ & $<00.1$ & - \\
\hline
Predicted $\mu_{SM}$ & $24.9\pm2.4$ & $20.2\pm1.7$ & $7.6\pm0.7$ & $5.85\pm0.6$ \\
Observed N$_{0}$    & 33 & 27 & 16 & 5 \\
$\mathcal{P}$(N$\ge$N$_{0}$/$\mu_{SM}$) & 9.3\% & 10.0\% & 0.7\% & 68.0\% \\ 
\hline
\hline 
\end{tabular}
\caption{The number $\mu_{SM}$ of events predicted by the SM, the number N$_{0}$
of  observed events and probability $\mathcal{P}$ that SM predictions  
fluctuate to no less then N$_{0}$ are presented. 
\label{tab:exp}
}
\end{minipage}
\end{tiny}
}}}
\end{table}
    
 
\hskip-0.6cm  
production of  LQ of first, second and third generation~\footnote{Since no evidence for the LQ 
production has been 
observed, all the results are reported as excluded limits at 95\% CL.}.  

D\O~ has  made  a search for  LQ pairs  decaying to 
$\nu\nu$+jets~\cite{D0-lepto}. The 
$\nu\nu$+jet candidate events are selected by requiring at least two jets with 
$\et>$ 50 GeV, 
$\met>$40 GeV, $\delta\phi(jet,\met)>$ 30$^{o}$, and jet-jet separation greater than
1.5~\footnote{Jet-jet separation is defined as $\sqrt{(\delta\eta)^{2}+(\delta\phi)^{2}}$, where
$\eta$ and $\phi$ are the jet pseudorapidity and azimuthal angle, respectively.}.
The main SM backgrounds are coming from the W and Z boson production associated with jets,
where the final states correspond to only neutrinos and jets, or to undetected charged lepton(s)  and jets,
or to an electron from W, which is misidentified as a jet, and an extra jet.    
The number of events surviving all of the selection cuts is 231( 242$\pm$18.9${^{+23.3}_{-19.0}}$ 
are expected from SM processes).  

A further step included a neural network optimization of the selection criteria for the production of 
100~GeV/c$^{2}$  scalar LQ and  for 200~GeV/c$^{2}$ vector LQ with  minimal vector coupling.
After applying the new criteria, 
58(10) events  for the scalar(vector) LQ are expected and $56^{+8.1}_{-8.2}$
($13.3^{+2.8}_{-2.6}$) are expected from the SM. This null result yields the 95\%~C.L.upper 
limit on the cross section (Fig.~\ref{fig:d0-lepto}) versus the leptoquark mass. LQ are  excluded 
with mass below 98~GeV/c$^{2}$  for scalar LQ, and  238~GeV/c$^{2}$ and  298~GeV/c$^{2}$ for 
vector LQ with minimal  vector coupling and  Yang-Mills coupling, respectively.     
 
All the current Tevatron LQ limits are summarized in Table~\ref{tab:lepto}.
    
\section{Conclusions}
The most recent results of the searches for new physics beyond the Standard Model
in the Tevatron Run~I data have shown some anomalies but no solid evidence for new
physics was found. 

CDF performed a detailed examination of the heavy flavor content of the jets produced 
with W bosons. An excess was found in events with a superjet, featuring both a SECVTX and SLT
tags in the same jet. In the  W+2,3 jet subsample, $4.4\pm0.6$ events are expected 
from the SM processes, while 13 are observed. By releasing some cuts and extending the search region 
4 more superjet
events were recovered  bringing to an effect of 
17 observed and  $4.8\pm0.7$ expected events.
A~ detailed~ examination of 
~the~ kinematical properties of the first~ 13 events shows~ that it  is 

\newpage
\begin{figure}
\begin{minipage}[p]{6.0cm}
\psfig{figure=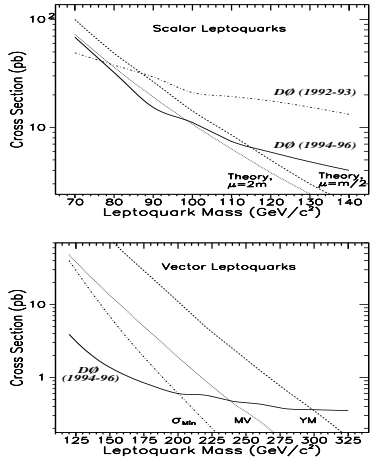,height=2.5in,width=2.0in}
\parbox[t]{2.0in}{\caption{
D\O~ limits on the cross section at 95\%~C.L. vs. LQ mass, for scalar (top) and vector (bottom) 
LQ. Different theoretical predictions are shown in the plots. 
\label{fig:d0-lepto}}}
\end{minipage}
\end{figure}

\hfill
\begin{table}
\vglue1truept\vskip-3.6in{\hglue1truept{\hskip2.6in{
\begin{tiny}
\begin{minipage}[p]{9.0cm}
\begin{tabular}{|cccc|}
\hline\hline 
          & Scalar    & Vector with & Vector with \\
BR(LQ$\rightarrow\ell^{\pm}$q) &            & Minimal Coupling & Yang-Mills Coupling\\
\hline
\multicolumn{4}{|c|}{First Generation LQ} \\
1.0 &  225 (220)   & 292(280) & 345(330)\\
0.5 &  204 (202)   & 282(265) & 337(310)\\
0.0 &   98         &  238     & 298     \\ 
\hline
\multicolumn{4}{|c|}{Second Generation LQ} \\
1.0 &  202 (208)   & 275(228) & 342(265)\\
0.5 &  170 (180)   & 270(230) & 310(290)\\
0.0 &   98         &  238     & 298     \\
\hline
\multicolumn{4}{|c|}{Third Generation LQ} \\
1.0 &  94 (148)   & 148(199) & 216(250) \\
0.0 &  98 (99)    & 238(170) & 298(225) \\
\hline
\hline 
\end{tabular}
\caption{Combined lover mass limits for LQ pair production for D\O~. The CDF limits are
presented in the brackets. 
\label{tab:lepto}}
\end{minipage}
\end{tiny}
}}}
\vspace{-0.2in}
\end{table}  

\hglue1truept{\hskip2.3in {
\parbox {9.4cm} {
statistically  difficult to reconcile them with a simulation of the
SM processes. The same SM simulation models well the 
complementary W+jet sample and other larger generic-jet samples of data.
There is not known  model which could incorporate the production and decay 
properties of these events. One is forced to conclude that either they are a rare fluctuation, 
or a hint for  something totally new. 
}}

\hskip-0.6cm         

Properties of the CDF events with photon and  lepton have been studied. An 
excess of events equivalent to 2.7 standard deviations has been found in one  subsample 
which features additional large $\met$.         
This is also an intriguing result which needs more data for confirmation.

Finally, we summarize the CDF and   D\O~ searches  for first, second and third 
generations  leptoquarks and  present the mass limits for  scalar and vector LQ's.   

Both CDF and D\O~ are  eagerly looking  forward to more data in the upcoming Fermilab Tevatron Run~II.

\vspace{-0.3cm}
\section*{Acknowledgments}
I would like to thank the organizers of this informative and enjoyable conference 
for their hospitality. We acknowledge the support of the US Department of Energy, the 
CDF and D\O~ collaboration  Institutions and their funding Agencies.


\vspace{-0.3cm}
\section*{References}
\begin{thebibliography}{99}
\bibitem{cdf}CDF Collaboration, D.~Amidei {\it et al}, \Journal{\NIM}{A271}{387}{1988}

\bibitem{d0}D\O~ Collaboration,  S.~Abachi {\it et al}, \Journal{\NIM}{A338}{185}{1994}.

\bibitem{discovery}CDF Collaboration, F.~Abe {\it et al}, \Journal{\PRD}{D50}{2966}{1994}.

\bibitem{massprd}CDF Collaboration, T.~Affolder {\it et al}, \Journal{\PRD}{D63}{032003}{2001}.

\bibitem{xsecprd}CDF Collaboration, T.~Affolder {\it et al}, \Journal{\PRD}{D64}{032002}{2001}.


\bibitem{superjet}CDF Collaboration, D.~Acosta {\it et al}, \Journal{\PRD}{D65}{052007}{2002}.

\bibitem{combined}G.~Apollinari {\it et al}, hep-ex/0109019,{2001}.

\bibitem{Kane}S.Ambrosanio{\it et al}, \Journal{\PRD}{D55}{1372}{1997}.

\bibitem{Allanach} B.C.Allanach, these proceedings.

\bibitem{gamma-lepton}CDF Collaboration, H.~Akimoto {\it et al}, hep-ex/0202044,(2002).

\bibitem{selection}CDF Collaboration, D.~Acosta {\it et al}, hep-ex/0110015,~(2001).

\bibitem{leptotheory-seeD0-1} H.~Georgi and S.~Glashow, \Journal{\PRL}{32}{438}{1974};\\
                              J.C.~Pati and A.~Salam,     \Journal{\PRD}{10}{275}{1974}
			     
\bibitem{scalar-seeD0-2} P.H.~Frampton,  \Journal{\MPLA}{7}{559}{1992}

\bibitem{leptotheory-seeD0-3} J.L.~Hewett and T.G.~Rizzo, \Journal{\PR}{183}{193}{1989}

\bibitem{D0-lepto} D\O~ Collaboration, V.M.~Abazov {\it et al}, hep-ex/0111047,(2001).

\end{thebibliography}
\end{document}